# COBE-DMR-NORMALISATION FOR COLD AND MIXED DARK MATTER MODELS

Krzysztof M. Górski[1,2,4], Radosław Stompor[3], and Anthony J. Banday[1]

## ABSTRACT

The two-year *COBE*-DMR 53 and 90 GHz sky maps are used to determine the normalisation of inflationary, flat, dark matter dominated universe models. The appropriately normalized cold and mixed dark matter models, computed for a range of values of $\Omega_b$ and $h$, and several values of the hot to cold dark matter abundance ratio, are then compared to various measures of structure in the universe.

*Subject headings:* cosmic microwave background — cosmology: observations — large-scale structure of the universe

Submitted to *Astrophysical Journal Letters*

[1] Universities Space Research Association, NASA/GSFC, Laboratory for Astronomy and Solar Physics, Code 685, Greenbelt, MD 20771

[2] Warsaw University Observatory, Aleje Ujazdowskie 4, 00-478 Warszawa, Poland

[3] Copernicus Astronomical Center, Bartycka 18, 00-716 Warszawa, Poland

[4] email: gorski@stars.gsfc.nasa.gov

## 1. Introduction

The *COBE*-DMR discovery of cosmic microwave background (CMB) anisotropy (Smoot et al. 1992, Bennett et al. 1992, Wright et al. 1992) has affected cosmology in both ontological and practical ways, but its predominant quantitative influence has been to provide the means for the accurate normalisation of theories of large scale structure formation.

The development of inflationary ideas during the 1980s induced a decade-long adherence to the cosmological paradigm which posits that the universe is spatially flat. Such a picture requires that the present energy density of the universe is dominated by non-baryonic dark matter or alternatively by a non-zero vacuum energy contribution (a cosmological constant term, $\Lambda$). The minimal version of the model, which invokes cold dark matter (CDM) as the major constituent of the universe, is presently in direct confrontation with astronomical observations. An extension of the model, which in addition to CDM postulates an admixture of hot dark matter (HDM), enjoys considerable popularity in contemporary cosmological research. Whilst vigorous discussion ensues in the literature as to the plausibility of the mixed dark matter (MDM) model as a viable cosmology, (see e.g. Schaefer, Schafi & Stecker 1989, Klypin et al. 1993, Pogosyan & Starobinsky 1993, Ma & Bertschinger 1994, Primack et al. 1994), we herein consider both CDM and MDM versions of the inflationary scenario (with $\Lambda = 0$).

In this *Letter* we use the linear angular power spectrum estimation technique (Górski 1994) to normalise the cold and mixed dark matter models to the two year *COBE*-DMR anisotropy measurements (Bennett et al . 1994), and subsequently discuss some predictions for large-scale structure measures resulting from this normalisation.†

## 2. CMB Anisotropy Normalisation Procedure

### 2.1 Data Selection and Power Spectrum Inference Method

The two year *COBE*-DMR 53 and 90 GHz sky maps are used identically as in Górski et al. (1994). In addition to the galactic sky maps we also utilise the ecliptic coordinate frame data sets as a check on the extent to which the coordinate dependent noise binning can affect the inferred normalisation.

We implement the power spectrum estimation method of Górski (1994). Coordinate system specific orthogonal basis functions for the Fourier decomposition of the sky maps are constructed so as to include exactly both pixelisation effects and a $|b| < 20°$ galactic plane excision (leaving 4016 and 4038 pixels in the galactic and ecliptic sky maps, respectively). A likelihood analysis is performed as detailed in Górski et al. (1994).

---

† Similar considerations were recently presented in Bunn et al. (1994).

## 2.2 Theoretical Spectra of Anisotropy

The cosmological models are specified to within an arbitrary amplitude as follows: 1) the global geometry is flat with $\Lambda = 0$ and $\Omega_0 = 1$, with the bulk mass density provided by either cold or mixed dark matter (CDM or MDM); for the MDM models the hot dark matter is introduced in the form of either one or two (equal mass) families of massive neutrinos, with the contributed fraction of critical density taken as $\Omega_\nu = 0.15$, 0.2, 0.25 and 0.3 for one massive flavour, and $\Omega_\nu = 0.2$, 0.3, otherwise; 2) for the CDM model the values of the Hubble constant, $H_0 = 100 \, h$ km s$^{-1}$ Mpc$^{-1}$, and baryon abundance are sampled at $h = 0.3$, 0.4, 0.5, 0.6, 0.7, and 0.8, and $\Omega_b = 0.01$, 0.03, 0.05, 0.07, 0.1 (for all values of $h$), respectively; in order to trace the Big-Bang nucleosynthesis (BBN) relation, $\Omega_b = 0.013 \, h^{-2}$ (Reeves 1994), we also use $\Omega_b = 0.02$ for $h = 0.8$, and $\Omega_b = 0.14$ for $h = 0.3$; for the MDM model $h = 0.5$ and $\Omega_b = 0.05$; 3) random-phase, Gaussian, scalar primordial curvature perturbations (no gravity waves) are assumed with the inflationary Harrison-Zel'dovich spectrum corresponding to an adiabatic density perturbation distribution, $P(k) \propto k$.

The CMB anisotropy multipole coefficients and the matter perturbation transfer functions for both models were evaluated using the Boltzman equation code of Stompor (1994) by solving the propagation equations up until the redshift $z = 0$. Over the low-$\ell$ range of CMB multipoles probed by $COBE$-DMR the theoretical spectra are indistinguishable for CDM and MDM models with equivalent $h$ and $\Omega_b$. Thus, the power spectrum amplitude derived from the data applies equally to both CDM and MDM models.

Figure 1 shows a selection of the CMB power spectra normalised to the $COBE$-DMR two year CMB anisotropy. We parameterize all such spectra by the exact value of the quadrupole, denoted $Q_{rms-PS}$, in a straightforward generalization of the $Q_{rms-PS}$ introduced in Smoot et al. (1992) for pure power law model spectra. This exactly evaluated present-day quadrupole is of smaller amplitude (by $\sim 10\%$, $5\%$, or $2\%$ for $h \lesssim 0.4$, $h \sim 0.5$, and $h \sim 0.8$, respectively) than its pure power law, Sachs & Wolfe (1967) counterpart, mainly due to the high-redshift, integrated Sachs-Wolfe effect.

## 2.3 Results of $Q_{rms-PS}$ Fitting

A typical likelihood fit of a flat dark matter model to the two year $COBE$-DMR data yields a $\sim 13\sigma$ significant determination of $Q_{rms-PS} \sim 20\,\mu K$. Systematic shifts in the central value of the fit are observed due to: 1) differences in the noise pixelisation in the galactic and ecliptic coordinate frames, which result in a $\sim 0.8\,\mu K$ difference between the inferred normalisation amplitudes, with higher values obtained from the ecliptic maps; 2) exclusion of the quadrupole, which produces a $\sim +0.4\,\mu K$ variation in the fitted amplitude; 3) the uncertainty in the values of $h$ and $\Omega_b$ (reflected by differences in spectral shape over the $\ell$-range accessible to DMR) causes an additional small spread, $\pm 0.25\,\mu K$, of the fitted amplitudes.

A convenient summation of the proposed overall normalisation for the flat, dark matter models is then $Q_{rms-PS} = (20 \pm 1.52)(\pm 0.4 \pm 0.2 \pm 0.25)\mu K$. The error ranges represent the statistical error and uncertainties associated with effects 1 through 3 above, respectively. It will be noted that the statistical error on the inferred normalisation is considerably larger than the other uncertainties.

Górski et al. (1994) showed that for the power law models specified by $Q_{rms-PS}$ and the spectral index $n$ a convenient, $n$-independent normalisation was in terms of the multipole amplitude $a_9 \simeq 8\,\mu K$. This $\ell$-order is related to the point at which the theoretical signal to noise ratio is of order unity. For the dark matter models discussed here, we find that the appropriate pivot point is at $a_{11} \simeq 7.15 \pm 0.55\mu K$.

### 3. Measures of Large-Scale Structure

Having determined the normalisations for our grid of CDM and MDM models, we now proceed to discuss the values of several large-scale structure statistics computed from the matter perturbation spectra according to the usual prescriptions given as footnotes to Table 1. Representative values for the $h = 0.5$, $\Omega_b = 0.05$ models are found therein.

### 3.1 Mass Fluctuations: $\sigma_8$ and $J_3$

$COBE$-DMR normalised values of $(\sigma_8)_{mass}$ are shown in Fig. 2a. A related, observable quantity is the fluctuation in the number of galaxies within a sphere of fixed radius. Recent estimates thereof are close to the standard value from Davis & Peebles (1983), $(\sigma_8)_{gal} \simeq 1$ — a representative selection of the galaxy surveys is: $(\sigma_8)_{gal} \simeq 0.8$ (Fisher et al. 1993), $0.87 \pm 0.07$ (Feldman, Kaiser & Peacock, 1994), and $0.83^{+0.05}_{-0.07}$ (Baugh & Efstathiou, 1993).

In order to compare the rms mass fluctuations with the rms galaxy fluctuations, we adopt the simple picture of biased galaxy formation, wherein $(\sigma_8)_{gal} = b\,(\sigma_8)_{mass}$, and $b$ is the linear bias factor. Some constraints on the value of $b$ can be imposed by recent galaxy surveys, but are more uncertain than the $(\sigma_8)_{gal}$ determinations. The estimated values of $b$ range from 0.9 to 2, with strong bimodal behaviour about $b \sim 1$ and 2 (see Table 1 in Dekel 1994, and Table 1 of Cole, Fisher & Weinberg 1994, with $\Omega_0 = 1$).

Estimates of $(\sigma_8)_{mass}$ inferred from galaxy cluster catalogues favour a higher value for $b$. Henry & Arnaud (1991) used the abundance of clusters as a function of X-ray temperature to derive $(\sigma_8)_{mass} = 0.59 \pm 0.02$ for a scale-free, flat universe. White, Efstathiou & Frenk (1993) have used the masses and abundances of rich clusters of galaxies to determine $(\sigma_8)_{mass} \simeq 0.52$ - 0.62 for a critical density universe (relatively independent of $h$). A low value of $(\sigma_8)_{mass} \sim 0.5$ is also required by the observed low pair-wise velocities (Davis et al. 1985, but see Zurek et al. 1994).

Although the value of the rms mass fluctuations is not yet well known, it seems unlikely that the biasing parameter is less than unity. This would require either pure CDM with a low-$h$ or high-$\Omega_b$, in disagreement with observations ($0.4 \lesssim h \lesssim 0.8$) or the BBN constraints, respectively, or alternatively an appreciable admixture of a hot dark component.

We have also computed the $J_3$-integral on a scale $R = 20h^{-1}\mathrm{Mpc}$ (Fig 2b). This, in principle, is less contaminated by non-linear galaxy power spectrum evolution than $\sigma_8$. The predictions of $J_3$ are compared to the value $J_3(20h^{-1}\mathrm{Mpc}) \simeq 700\,h^{-3}\,\mathrm{Mpc}^3$ from Davis & Peebles (1983).

3.2 Galaxy and Matter Distribution Power Spectra

In principle, the most direct and informative comparison of theory and observations can be conducted using the predicted and empirical power spectra. Several recently estimated galaxy power spectra are shown in Fig.3, together with various theoretical linear power spectra (assuming $b = 1$). However, such a comparison is still subject to difficulties. There are notable discrepancies in both shape and amplitude between galaxy power spectra derived by varying techniques from different samples. This may be due to the different power spectrum estimators employed, or intrinsic variations in the samples. Furthermore, on small scales, the effects of non-linear evolution should be apparent, and possible distortions can be introduced by redshift space mapping.

With these caveats, we only proceed to use Fig. 3 illustratively. It is apparent that none of the empirical power spectra exhibit as much small-scale power as the linear *COBE*-DMR normalised CDM spectrum †. This could be considered as supportive of the MDM models. However, the high amplitude of the matter power spectra implied by the *COBE*-DMR measurements leaves very little room for the biasing parameter to exceed unity, in conflict with the $b$-estimates from cluster properties. Baugh & Efstathiou (1993) have suggested that, with $b \sim 2$, the matter power spectrum required by the observed galaxy distribution should look rather like that for the $\Omega_0 = 1$, $\Gamma = 0.2$ model ‡, depicted in Fig. 3 as a thin solid line. Alas, *none* of the *COBE*-DMR-normalised CDM or MDM model power spectra resembles such an *ad hoc* spectrum.

### 3.3 Large-Scale Flows

The local streaming motions of galaxies provide an interesting constraint for cosmological models. In particular, galaxy peculiar velocities directly measure mass fluctuations, independently of a linear bias parameter. Dekel (1994) gives estimates of the average peculiar velocities within spheres of radius 1000 to 6000 km s$^{-1}$. These seem to be in agreement with the HI data from Giovanelli & Haynes (Dekel, private communication). An important recent development in the field was the determination by Lauer & Postman (1994) of significant bulk flow in a deep volume limited sample of $\sim 100$ galaxy clusters. If confirmed, this observation would likely invalidate all of the presently considered models of structure formation, which predict too rapid a decrease in the bulk flow amplitude with scale. The *COBE*-DMR normalised rms bulk flows are shown in Fig. 2c. The model predictions are in good agreement with the POTENT data, while the disagreement with the Lauer & Postman result appears rather significant.

---

† This conclusion is based on the real space power spectrum estimates from Baugh & Efstathiou (1993) and Peacock & Dodds (1994), and the redshift space power spectra derived from the IRAS 1.2 Jy survey. However, the redshift space spectra from da Costa *et al.* (1994) indicate yet more power on small scales, a discrepancy which can only become more pronounced when one recalls that the real-to-redshift space mapping suppresses power on scales $k \gtrsim 0.1h\mathrm{Mpc}^{-1}$ (Gramann *et al.* 1993). Similarly, the predicted linear real-to-redshift space power enhancement on scales $k \lesssim 0.03h\mathrm{Mpc}^{-1}$) is too small to account for the amplitude discrepancies in these data sets for $b \gtrsim 0.4$.

‡ Here, we follow the parametrization of the matter power spectra introduced by Efstathiou, Bond, & White 1991.

## 4. Discussion

The improved quality of the two year $COBE$-DMR data combined with reliable power spectrum estimation techniques allows the accurate normalisation of cosmological theories. Previously, the variance of $COBE$-DMR temperature fluctuations on a $10°$ angular scale was utilised for the normalisation of the power spectrum. Subsequent work (Wright et al. 1994, Banday et al. 1994) has demonstrated that this technique can be unreliable without considerable attention. More appropriate methods take full advantage of the measured CMB anisotropy power distribution on all angular scales accessible to the $COBE$-DMR instrument, as implemented in this $Letter$.

The cold dark matter theory with a standard choice of cosmological parameters requires a very high normalisation in order to fit the CMB anisotropy distribution. Analysis of the first year of $COBE$-DMR data had already suggested that $(\sigma_8)_{mass} \sim 1$ (Wright et al. 1992, Efstathiou, Bond, & White 1992), and this value increases to $\sim 1.4$ with two years of data and an improved analysis technique. Although this normalisation allows the theory to predict large-scale velocities of comfortably high amplitude, it also results in a significant overproduction of density perturbations on scales of $\lesssim 20\,h^{-1}\mathrm{Mpc}$. CDM has often been criticised for its poor match to both galaxy and cluster distributions. Mixed dark matter models manage to circumvent, to a certain degree, these same problems by construction — massive neutrinos partially damp the perturbations at those length-scales where CDM looks problematic. Among the MDM models those with two species of massive neutrino seem to meet the observational constraints more comfortably (see also Primack et al. 1994). The larger free-streaming radius allows for the suppression of the perturbation amplitude on larger scales than in models with one massive flavour. This is reflected in the decrease of the predicted $(\sigma_8)_{mass}$ values. Nevertheless, the proponents of MDM will have to address the viability of the model viz. the simultaneous requirements that there should be no bias between the galaxy and mass distribution (as suggested by this analysis) and the galaxy pair-wise velocities should be small.

Whilst it would be safer to await the final 4-year COBE results before offering definitive statements as to the viability of theoretical models, one should note that the CDM normalisation derived from the two year $COBE$-DMR data does appear to be irreconcilably high, while the MDM model has little room left for adjustment.

We acknowledge the efforts of those contributing to the $COBE$-DMR. This work was supported in part by the Office of Space Sciences of NASA Headquarters. We are grateful to L. da Costa, K. Fisher, M. Vogeley for providing their power spectra, and A. Dekel for providing the bulk flow points.

# FIGURE CAPTIONS

Fig. 1.— CDM/MDM CMB anisotropy power spectra normalised to $COBE$-DMR (where $\Delta_\ell^2 = \ell(2\ell+1)a_\ell^2/4\pi$). The four sets of curves correspond to the power spectrum amplitude determination as follows (from top to bottom): 1) ecliptic coordinates, quadrupole excluded, 2) ecliptic coordinates, quadrupole included, 3) galactic coordinates, quadrupole excluded, and 4) galactic coordinates, quadrupole included. Within each set the three curves denote: $h = 0.5$ and $\Omega_b = 0.05$ - heavy type, $h = 0.3$ and $\Omega_b = 0.14$ - medium-heavy type, $h = 0.8$ and $\Omega_b = 0.02$ - light type, which are consistent with the constraints from BBN. The crossing point within each set is at $\ell \approx 11$. Several pure Sachs-Wolfe power law spectra are shown (shifted down to 0.9 at $\ell = 2$). Although a power law approximation, $P(k) \propto k^n$, used to generate the multipole coefficients $a_\ell^2$ solely through the Sachs & Wolfe (1967) effect will be poor for such spectra, a value of $n \lesssim 1.1$ would be most appropriate over the range $\ell \lesssim 15$. This is a little steeper than the underlying, inflationary $n = 1$ spectrum (Bond 1993). The proposed overall two year $COBE$-DMR normalisation (translated from $Q_{rms-PS} \sim 20\,\mu$K) is represented by a filled circle; the error bar represents a typical statistical uncertainty of an individual likelihood fit. Clearly, this encompasses the fit uncertainties due to choice of coordinate system, quadrupole inclusion/exclusion, and/or cosmological parameter values.

Fig. 2.— a) $(\sigma_8)_{mass}$ values predicted from the $COBE$-DMR normalised, flat, dark matter dominated models. The thin solid lines correspond to the CDM model with $h$ decreasing from 0.8 to 0.3 in steps of 0.1 from top to bottom. The thick solid line shows the models which obey the Big Bang nucleosynthesis constraint, $\Omega_b = 0.013\,h^{-2}$. The individual points represent several mixed dark matter models: triangle - $\Omega_\nu = 0.15$, square - $\Omega_\nu = 0.20$, hexagon - $\Omega_\nu = 0.25$, diamond - $\Omega_\nu = 0.30$, where the filled symbols have $N_\nu = 1$, open symbols $N_\nu = 2$. All of the points correspond to $h = 0.5$ and $\Omega_b = 0.05$ but are spread out on the plot for clarity.

b) The values for the $J_3$ integral over the density perturbation correlation function within $20\,h^{-1}$Mpc (units $[h^{-1}\mathrm{Mpc}]^3$). Same coding as a.

c) rms amplitudes of the large scale flows ($[\mathrm{km\,s^{-1}}]$). The heavy lines correspond to the CDM models with $h = 0.8$, $\Omega_b = 0.02$ — top, $h = 0.5$, $\Omega_b = 0.05$ — middle, $h = 0.3$, $\Omega_b = 0.14$ — bottom. Thin lines separating from the $h = 0.5$ curve correspond to the MDM models with $N_\nu = 1$ (higher line), and $N_\nu = 2$, practically independent of $\Omega_\nu$.

POTENT data (courtesy A. Dekel) are shown by circles, and the square shows the Lauer & Postman datum.

Fig. 3.— *COBE*-DMR normalised inhomogeneity power spectra and, shown as the lower-most curves, miscellaneous spectral windows required for the computation of the statistics considered in this paper (as annotated). The theoretical mass distribution power spectra for the $\Omega_0 = 1$, $h = 0.5$, $\Omega_b = 0.05$ models are shown in the middle section of the plot: the top heavy line shows the CDM power spectrum; the lower heavy lines correspond to two $N_\nu = 1$ MDM models — $\Omega_\nu = 0.2$ (upper) and $\Omega_\nu = 0.3$ (lower); the medium heavy lines correspond to the equivalent $N_\nu = 2$ MDM models. Note that the MDM transfer function for $N_\nu = 2$ drops faster near $k \sim 0.04$ than for $N_\nu = 1$. The thin line shows the $\Gamma = 0.2$ transfer function for CDM. The vertical error bar on the $P(k) \sim k$ part of the spectrum above the $a_{11}$ window illustrates the $\sim 1\sigma$ allowed variation in amplitude for this fixed slope. Conversely, the superposed 'bow' shows the allowed $1\sigma$-variation ($\sim 0.3$) in tilt for spectra with the fixed $a_{11} = 7.15\,\mu$K amplitude. It should be noted that this is the proper representation of two dimensional uncertainty in the spectrum determination from the two year *COBE*-DMR data for flat dark matter models. One is *not* at liberty to arbitrarily vary both amplitude and spectral slope simultaneously. This is equivalent to the observed degeneracy seen in the two-dimensional ($Q_{rms-PS}$, $n$) fits to pure power law models (Górski et al. 1994 and references therein). The upper section of the plot reproduces the theoretical spectra for CDM, MDM ($N_\nu = 1$, $\Omega_\nu = 0.2$), and $\Gamma = 0.2$-CDM – all shifted upward for clarity. Overplotted are several observational estimates for the galaxy distribution power spectrum: squares - Baugh & Efstathiou (1993), filled circles - Peacock & Dodds (1994), pentagons - Fisher et al. (1993), triangles - da Costa et al. (1994).

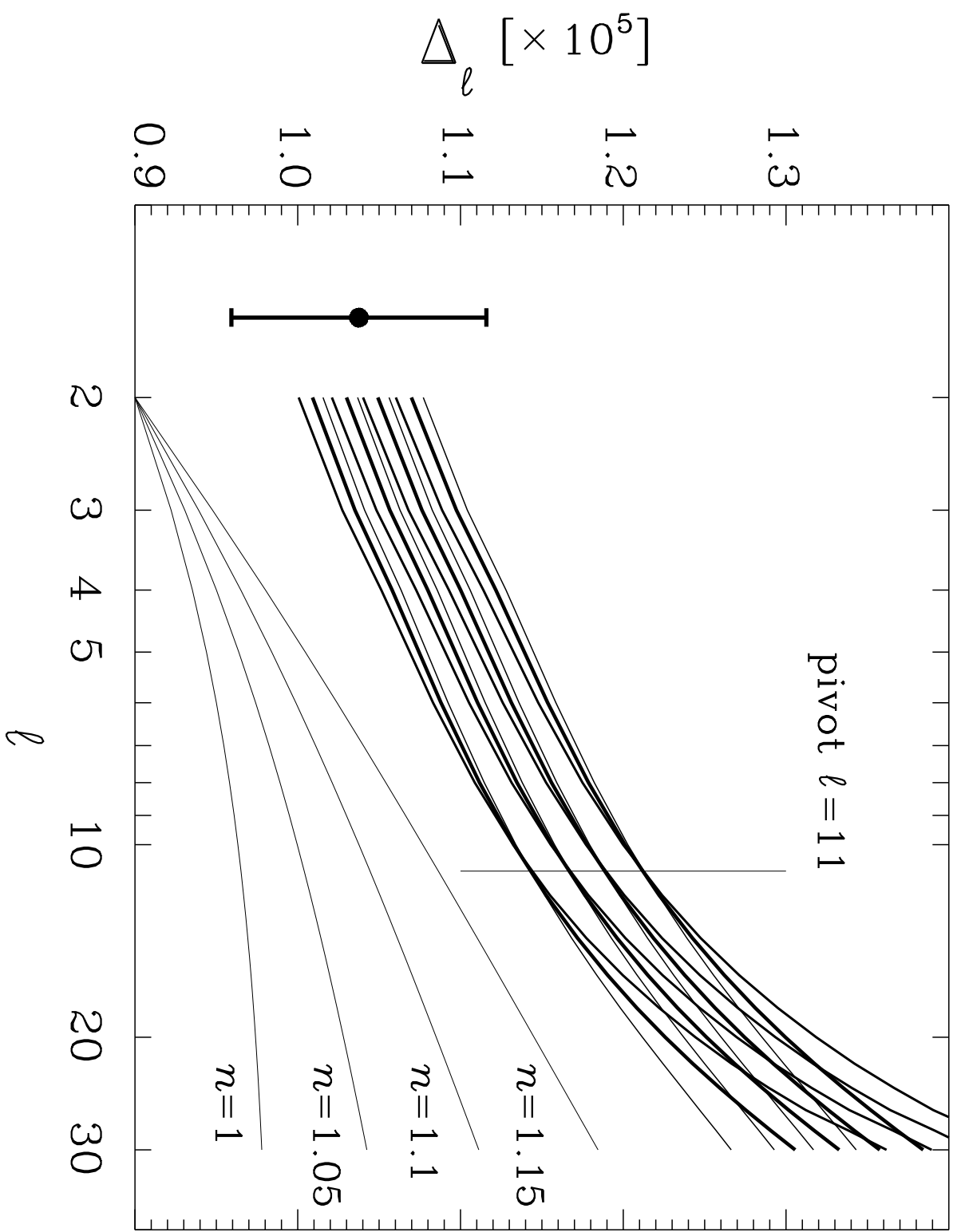

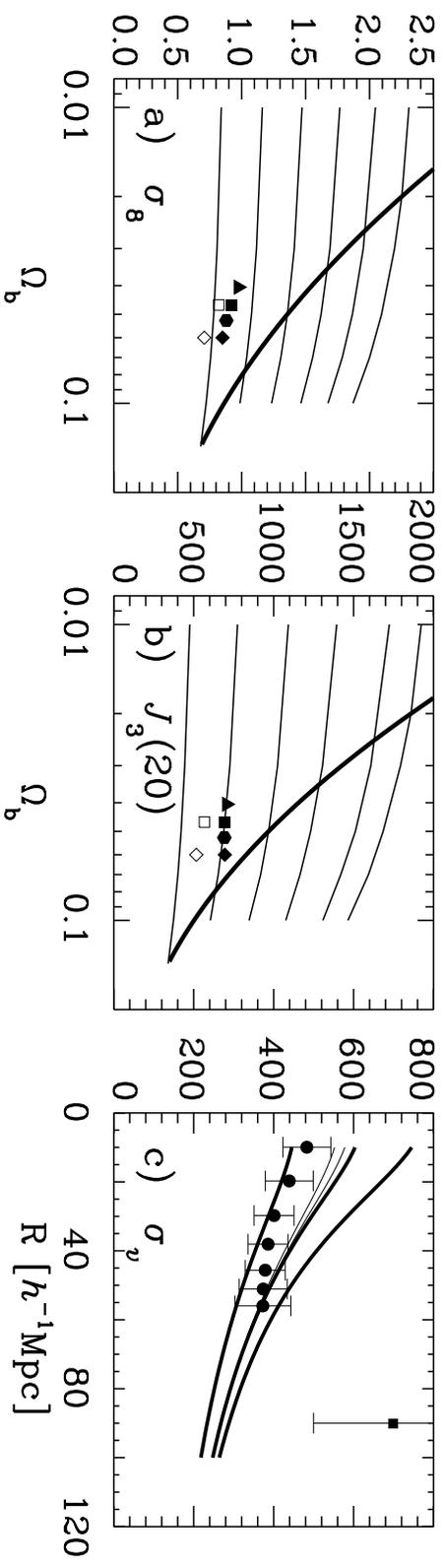

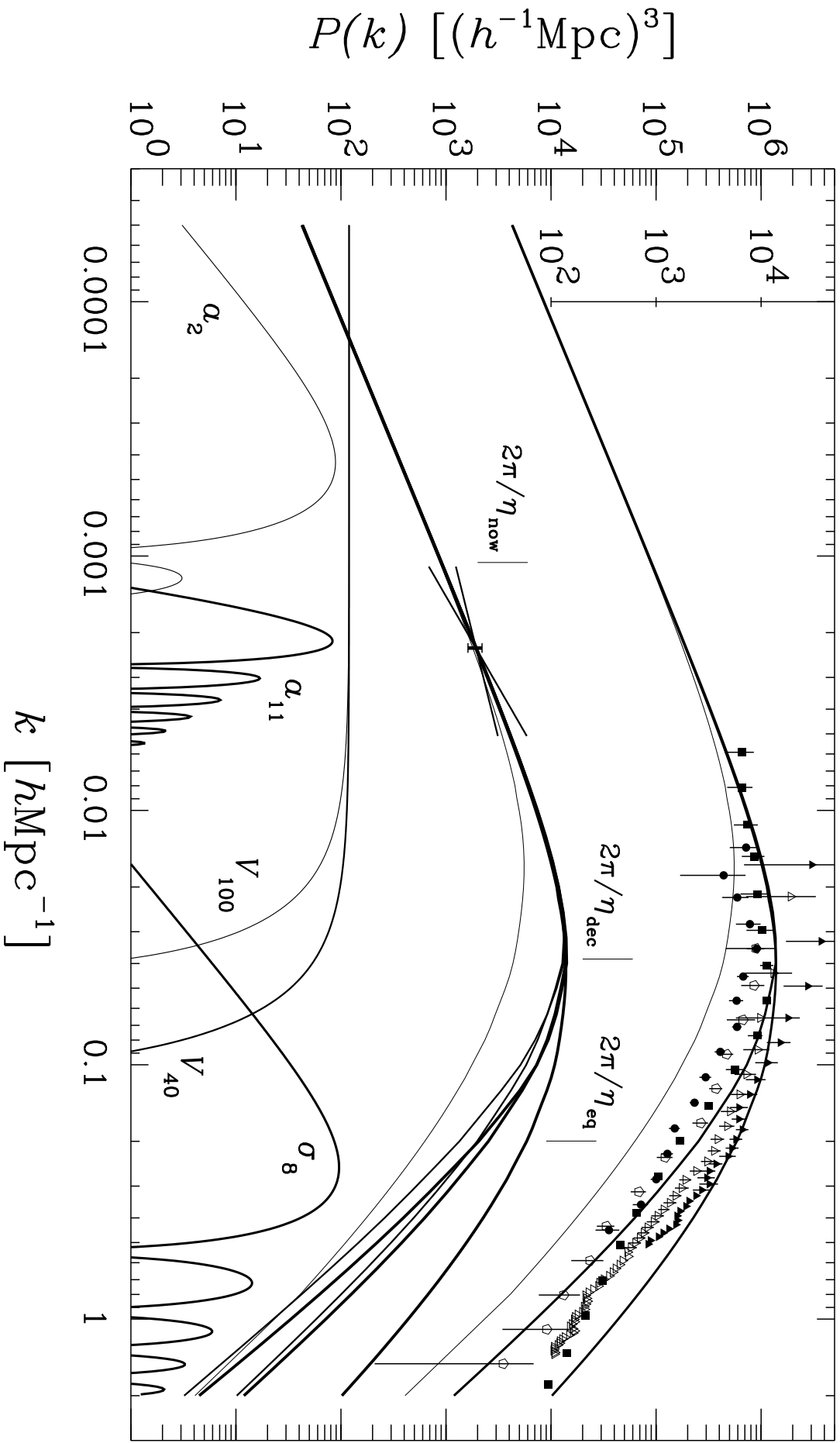

Table 1: Inferred cosmological statistics for models with $\Omega_0 = \Omega_{CDM} + \Omega_\nu + \Omega_b = 1$, $\Omega_b = 0.05$, $h = 0.5$, and a $COBE$-DMR normalisation of $Q_{rms-PS} = 20.04$ $\mu$K. $N_\nu$ is the number of massive neutrino species, and $m_\nu$ the neutrino mass in eV. The errors, including both statistical ($1\sigma$) and systematic deviations, are of the order of 11%.

| $\Omega_\nu$ | $N_\nu$ | $m_\nu$ | $(\sigma_8)^{(a)}_{mass}$ | $J_3(20)^{(b)}$ | $v^{(c)}_{40}$ | $v^{(c)}_{60}$ | $v^{(c)}_{100}$ |
|---|---|---|---|---|---|---|---|
| 0.00 | – | 0.0 | 1.36 | 968 | 444 | 355 | 248 |
| 0.15 | 1 | 3.7 | 0.97 | 706 | 441 | 357 | 251 |
| 0.20 | 1 | 4.9 | 0.92 | 694 | 442 | 358 | 252 |
| 0.25 | 1 | 6.1 | 0.88 | 691 | 444 | 359 | 252 |
| 0.30 | 1 | 7.3 | 0.85 | 695 | 445 | 360 | 252 |
| 0.20 | 2 | 2.4 | 0.82 | 567 | 435 | 356 | 252 |
| 0.30 | 2 | 3.7 | 0.71 | 516 | 439 | 359 | 254 |

$^{(a)}$ $(\sigma_{hR})^2_{mass} = \frac{1}{2\pi^2} \int_0^\infty w^2_{TH}(kR) \, P(k) k^2 \, dk$

$^{(b)}$ $J_3(hR) = \frac{R^3}{2\pi^2} \int_0^\infty w^2_{TH}(kR) \, P(k) \, k^2 \, dk$

$^{(c)}$ $v^2_{hR} = \frac{H_0^2}{2\pi^2} \int_0^\infty w^2_{TH}(kR) \, e^{-k^2 r_*^2} \, P(k) \, dk$, $hr_s = 12$Mpc
$w_{TH}(x) = 3j_1(x)/x$